# Enabling Zero Trust Security in IoMT Edge Network

Maha Ali Allouzi
mallouzi@kent.edu
Javed Khan
javed@kent.edu

**Abstract:**

Internet of Medical Things (IoMT) deals with a patient-data-rich segment, which makes security and privacy a severe concern for patients. Therefore, access control is a significant aspect of ensuring trust in the IoMT. However, deploying existing authentication and authorization solutions to the Internet of Medical Things (IoMT) is not straightforward because of highly dynamic and possibly unprotected environments and untrusted supply chain for the IoT devices. In this article, we propose *Soter*, a Zero-Trust based authentication system for the IoMT. Soter Incorporates trust negotiation mechanisms within the Zero Trust framework to enable dynamic trust establishment. When a user or device seeks access to a resource, initiate a trust negotiation process. During this process, credentials, attributes, and contextual information are exchanged between the requester and the resource owner. *Soter* defines access rules based on various factors, including user identity, device health, and location. Access is granted or denied based on these conditions.

## I. INTRODUCTION

The Internet of Medical Things (IoMT) is an evolving technology intended to improve patient's quality of life by enabling personalized e-health services without time and location constraints. However, IoMT devices (e.g., medical sensors and actuators) that compose the critical fundamental elements of the IoMT edge network and form what is referred to as a wireless Body Area Network (WBAN) are vulnerable to various types of security threats, and thus, they cause a significant risk to patient's privacy and safety [1]. A 2022 report by the FBI found that 53% of digital medical devices and other Internet-connected devices have at least one critical vulnerability that remains unpatched [1]. Based on that and the fact that security and privacy are essential factors for the successful adaptation of IoMT technology into pervasive healthcare systems, there is a severe need for novel security mechanisms to preserve the security of the IoMT edge network (WBAN) as utilizing existing authentication solutions to the Internet of Medical Things (IoMT) is not straightforward because of highly dynamic and possibly unprotected environments, and untrusted supply chain for the IoT devices.

To overcome this issue, *Soter* uses Zero Trust Management (ZTM) as an authentication and authorization mechanism; ZTM is essentially an access control in a widely distributed environment where authorization cannot be based on identity authentication using Credential service providers (CSPs). So Soter will Figure 1 shows Soter's basic protocol flow.

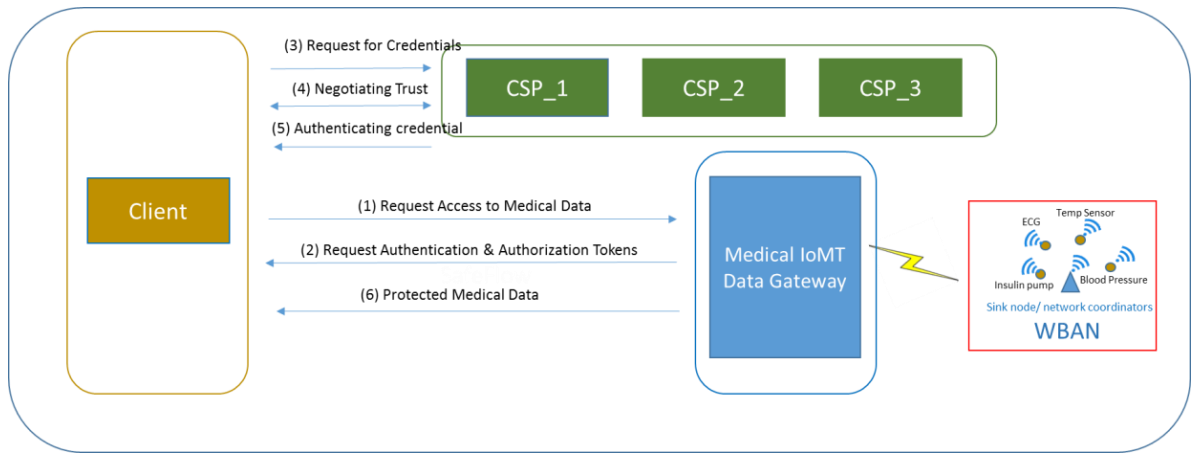

*Figure 1 Soter basic Protocol flow*

In this work, we evaluate, adapt, and implement an access control system based on Zero Trust Management such that it is resilient to the characteristics and threats of IoMT environments, which currently has not been addressed in existing access control systems used in IoT environments. Section II presents related work. Section III identifies gaps between existing access control systems in IoMT environments. Section VI presents the threat model that guided the architecture and implementation of Soter. Section V presents the evaluation of Soter. Finally, section VII presents conclusions and future works.

## II. BACKGROUND AND RELATED WORK

Access control theory has been the focus of academic and industry research in recent decades. A combination of theory and software engineering techniques has been developed and deployed on systems in the past few years. In this section, we introduce the problem of modeling dynamic access control systems based on Zero Trust and review some relevant related work. Implementation of the access control system can be divided into three design layers:

- **Access Control Policy**: This is a high-level description of conditions and rules under which the requestor can access the protected resource. Several policy languages have been developed and used in access control systems; however, we could not adapt any in Soter as they don't support multiparty part trusts and are unsuitable for IoMT constraint devices. So, we designed MedDL [2] for Soter, a Role and policy-based language that supports credential-based access control in cross-domain security.
- **Access Control Model**: formal presentation of how policy is enforced; it is a link between policy and mechanism. Some well-known models are Discretionary access control (DAC), Mandatory access control (MAC), Role-based access control (RBAC), and Attribute-based access control (ABAC).
- **Access Control Mechanism**: is a policy enforcement mechanism using the system's acceptable structure.

The way the access control systems were first designed did not consider the heterogeneity and the openness of today's IoMT systems, which made these access control systems unsuitable for use in such systems. Therefore, Authentication and Authorization in IoT environments have received much attention in recent years. [3] Designed OSCAR, an approach for access control in the IoT using object security based on secret keys for authorization to access resources. However, this methodology is not scalable as fine-grained access rights require managing an increasing number of secret keys, which is difficult in open, decentralized environments such as the IoMT. [4] Propose AoT, an authentication and access control scheme for the IoT device life cycle. AoT uses Identity- and Attribute based Cryptography with ABAC. Their results show that this method does not apply to the environment of constrained devices. [5] Propose OAuth-IoT, an access control system that adapts ACE, assuming that ACE would not work if the client were outside the IoT, which is a flawed assumption. Finally, [6] proposes a framework for authentication and authorization for IoT devices in disadvantaged environments. Their work requires the client to preregister with the authorization server and is based on static policies agreed upon before the authentication happens. Parties might need to change their authentication requirements and policies, and they might want to have a dynamic access control policy. Therefore, in Soter, we build our authentication using Zero Trust Management. This credential-based trust negotiation system enables parties to have a fixable and dynamic policy rule that can be changed. The operation of Soter will be discussed in section V. Table I lists the notations and the terms used throughout this paper.

Table I Notations and Terms used in this paper.

| Notation | Description |
|---|---|
| $C_j^i$ | Credential $i$ for Device $j$ |
| CSP | Credential Service Provider, which is an Identity provider (IdP) |
| Client | The requestor of the medical information |
| Resource Server | The provider of the medical data |
| IoMT Edge Network | The wireless body area network is composed of IoMT devices. |
| DODAG | Destination Oriented Directed Acyclic Graph, RFC 6550 |
| DIO | DODAG Information Object, RFC 6550 |
| DAO | Destination advertising object, RFC 6550 |
| CoAP | The constrained Application protocol, RFC 7250 |
| CBOR | Concise Binary Object Representation, RFC 7049 |
| COSE | CBOR object Signing and Encryption, RFC 8152 |
| DTLS | Datagram Transport Layer Security, RFC 6347 |
| $A\_C_1I_1R_1$ | Active communication mode involves one client, CSP, and RS, responsible for collecting the required credentials. |
| $A\_C_1I_2R_1$ | Active communication mode involves one client, two CSPs, and one RS, which is responsible for collecting the required credentials. |
| $A\_C_1I_3R_1$ | Active communication mode involves one client, three CSPs, and one RS responsible for collecting the required credentials. |
| $I\_C_1I_1R_1$ | Inactive communication mode involves one client, CSP, and RS; the client collects the required credentials. |
| $I\_C_1I_2R_1$ | Inactive communication mode involves one client, two CSPs, and one RS; the client is responsible for collecting the required credentials. |
| $I\_C_1I_3R_1$ | Inactive communication mode involves one client, three CSPs, and one RS; the client is responsible for collecting the required credentials. |

### III. USING ZERO TRUST MANAGEMENT IN IOMT ENVIRONMENTS

Most existing access control mechanisms for authentication and authorization assume that Clients and Resource Servers have securely exchanged their credentials with the Credential Service Providers (CSP) beforehand. This implies that neither the client nor the Resource Server has any control over what credentials they are willing to disclose, which means they cannot manage the rules governing the release of their credentials. The exchange of credentials takes place before the authorization phase. Other mechanisms rely on secret keys for authorization to access resources, but this approach is not scalable. Fine-grained access rights require managing more secret keys, which is difficult in open, decentralized environments such as the Internet of Medical Things (IoMT) [3]. To overcome these two gaps, Soter based its authentication and authorization on Zero Trust Management, where parties build trust by exchanging credentials based on dynamic policy rules that govern the release of these credentials; these rules are flexible and can be modified when needed with no system change requirements. Zero Trust Management is suitable for the IoMT environment as no prior registration and cryptographic key management are required. To the best of our knowledge, Soter is the first access control system to use the TM for authentication and authorization in the IoMT environment. Section VI illustrates and describes the components and the architecture of Soter.

### IV. SOTER THREAT MODEL

The threat model for Soter is based on its participants' entities: Client, one or more Credential Service Providers, and one Resource Server. Using the Microsoft Threat Modeling Tool [7], we developed separate diagrams for Soter's different modes of operations. Microsoft threat tool applies the STRIDE threat model [8] to each element in the diagram (process, dataflow, and data store) and provides a guided analysis of the threats and mitigations. The first mode is called active communication mode, where the Resource Server is an active entity responsible for collecting all the credentials, which are digitally signed certificates asserting attributes about their subjects and used as the basis of access control decisions in Soter. We have designed different scenarios for this mode, including one or more CSPs. The first scenario, A_C1I1R1, indicates Active mode with one client, CSP, and RS, as shown in Figure 2. The tool identified 12 threats based on the security requirement violated, as revealed in Table II. For the Inactive mode shown in Figure 3, the RS is inactive, and the client is responsible for collecting the required credentials; the tool identified 18 threats. Figure 4 shows our threat analysis results for Soter systems with multiple numbers of CSPs included in the access control decision.

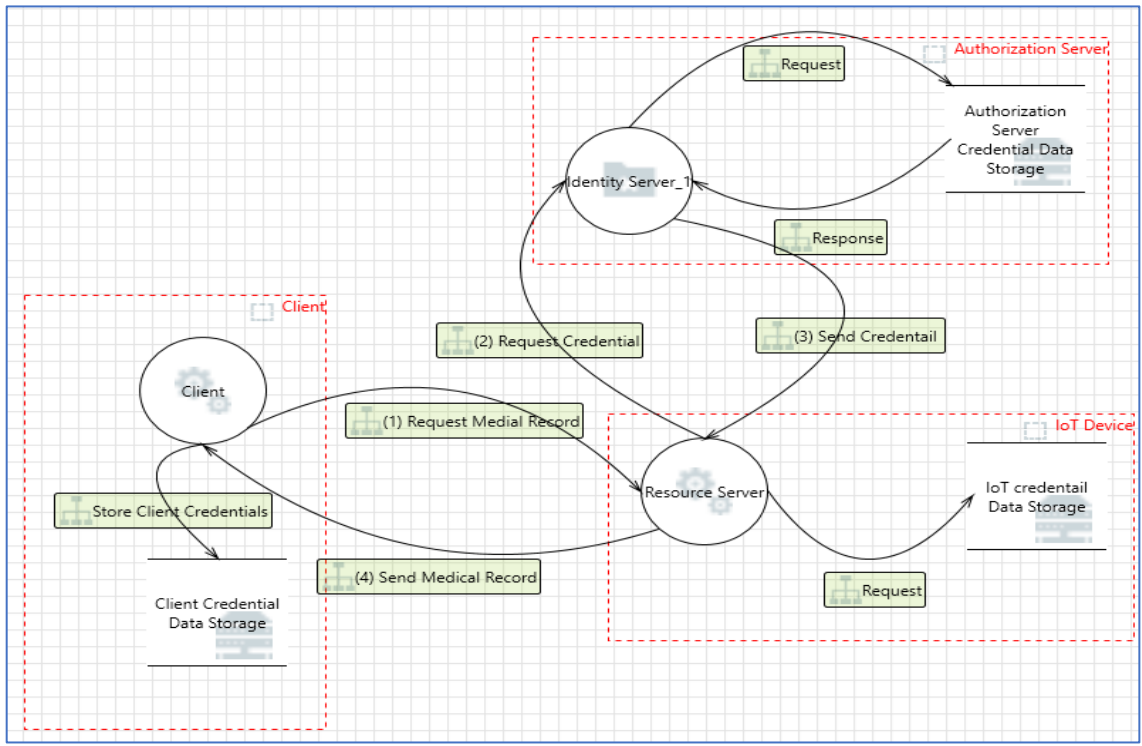

*Figure 2 STRIDE Data Flow Diagram Active_$C_1I_1R_1$*

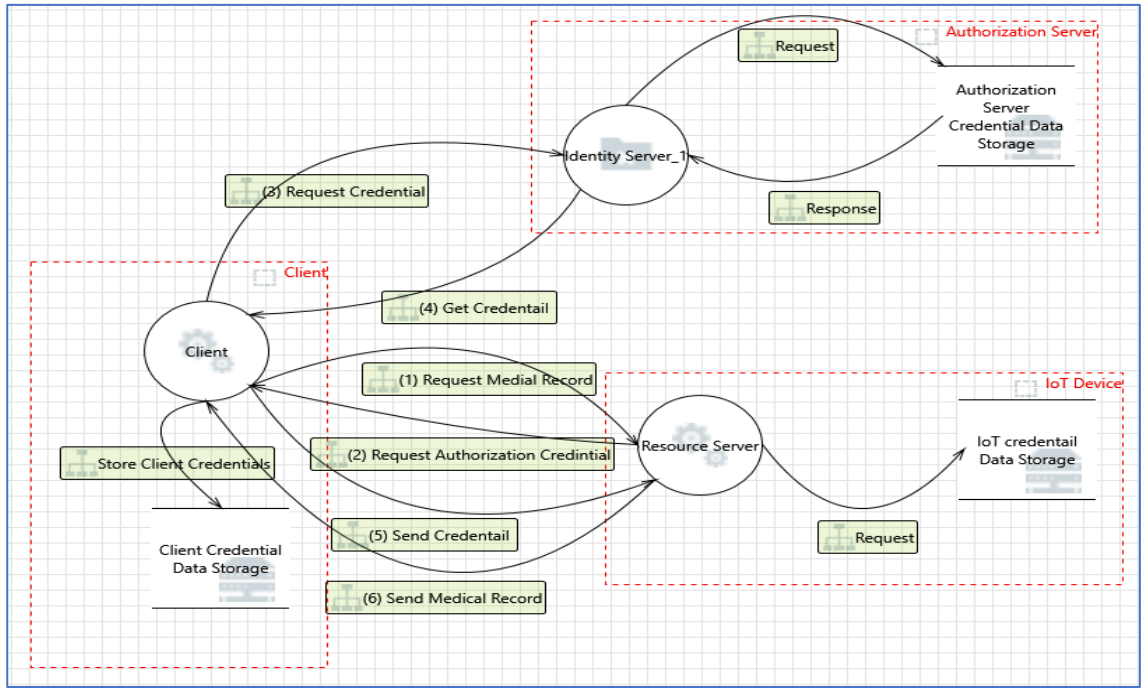

Figure 3 STRIDE Data Flow Diagram Inactive _$C_1I_1R_1$

Table II Threat Model Results

| CATEGORY | NAME OF THE ATTACK | DESCRIPTION | EXAMPLE |
|---|---|---|---|
| Spoofing | Node impersonate | adversary can impersonate CLIENT to gain unauthorized access to CSP and obtain credential data | An attacker can impersonate the doctor laptop [client] to gain unauthorized access to the Hospital Employee's DB [CSP] and obtain the doctor medical ID [credentials] that can be used to access [Medical Data resources] |
| | | adversary can impersonate CLIENT to gain unauthorized access to Medical Resource Server and obtain medical data | An attacker can impersonate the doctor laptop [client] to gain unauthorized access to Medical Records' DB [Medical Data] to read/write/modify Patients medical records |
| Information Disclosure | Sniffing data flows | An adversary may issue valid tokens if Identity server's signing keys are compromised | An attacker can sniff data flows between Client and the CSP and between CSP and Medical Data Gateway to obtain credential tokens that can be used to access medical data gateway server. |
| Tampering | Changing Data flows | An Adversary can temper data flow to obtain unauthorized access to nodes | An attacker can change the message sent from the Resource Server to the Client and get access to the Medical Records |
| Information Disclosure | Unauthorized access to credential data stores | An adversary may gain access to credential data stores on any node during the trust negotiation and use them to access other nodes | An attacker can change the message sent from the Resource Server to the Client and get access to the Medical Records |
| Elevation of Privileges | performing of operations other than that authorized operation | Malicious nodes( Client or Medical Data Resource server can obtain access to run other code to compromise the CSP | An attacker can run other code on the CSP and obtain credential or compromised the CSP ( viruses and worms) |

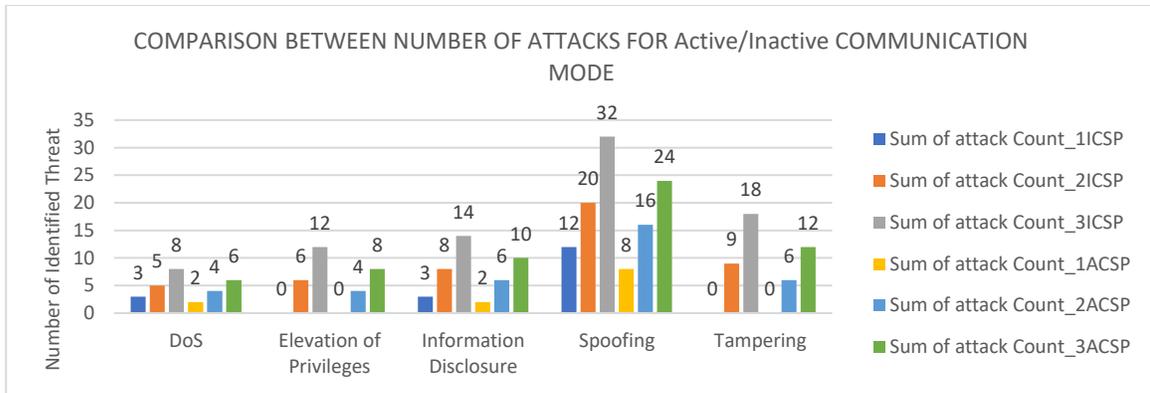

*Figure 4 Comparison between the number of attacks for Active and Inactive communication mode in Soter*

## V. SYSTEM ARCHITECTURE

In this section, we describe Soter's architecture, design, and implementation that addresses the limitations identified in section III and the threats presented in section IV. The design is divided into two phases, Phase I and II, as shown in Figure 5.

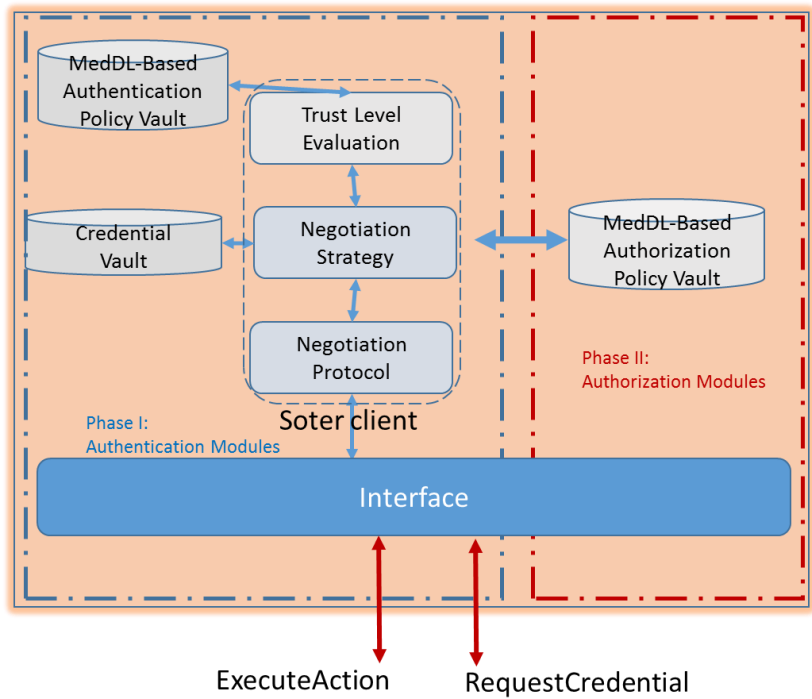

*Figure 5 Soter Component Architecture*

***Phase I*** will be initiated when the client requests to execute an action on the IoMT devices, like reading medical records, starting reading a health parameter, etc. To authorize this action, Soter should first authenticate the client and then see if this authenticated client is authorized to execute this action. As stated before, Zero Trust Management is used to authenticate clients. Following is an explanation of the role each module has:

- *Negotiation Protocol*: An algorithm that orchestrates and manages a negotiation session, how to start a negotiation, when it is a particular peer's turn to send a message, what the message formats are, and how to tell whether a negotiation succeeds or fails. The protocol depends on the resolution algorithm to decide this information.
- *Negotiation Resolution Strategy*: An algorithm that decides the content of each negotiation message, what credentials to release, and what to request from other peers.
- **Trust Level Evaluation**: This module distinguishes the different levels of trustworthiness the client can belong to; this module is added to reduce the number of exchanged credentials for higher trust level clients.
- **Authentication Policy vault**: A set of policies that permits or denies access to member's credentials.
- *Credential Vault*: A set of Digital documents or certificates serving as official proof of the holder's identity or attributes. The notation $C_j^i$ was used to represent member i's j<sup>th</sup> credentials.
- *Negotiation Message:* This contains all necessary state information about credentials involved in the negotiation, their states, values, and owners, as well as the state of the negotiation process and the session ID required when peers are involved in more than one concurrent negotiation process.

During Phase II, authenticated clients requesting to execute actions on the IoMT edge network should be authorized; This is done based on the Authorization Policy vault that has the rules governing the authorization of actions and requests sent to the resource server. The authorization policy vault has all the rules written using our MedDL [2] a Datalog with constraint policy language, with the HIPAA and the IoMT as the constraint domains.

## VI. SOTER IMPLEMENTATION

A dynamic view of the SOTER system is shown in Figure 6. Three main components correspond to the three SOTER participated entities: Client, RS, and one or more CSP.

We used Python to implement three sets of libraries, all of which implement specific protocols and entities required to run the Soter protocol flow.

**Soter Library**

The Soter library consists of implementations for all three Soter entities, i.e., the Client, the Resource server, and the Credential Server Provider.

**COSE Library**

The COSE library models object from the COSE standard; This includes CBOR encode encryption objects, digital signature objects, and COSE formatted keys.

**DTLS Library**

The DTLS library is used to secure the CoAP protocol. DTLS is an adaptation of TLS designed to use UDP as its transport protocol instead of TCP. DTLS protocol reduces the overhead and size of the message significantly.

The source code and a small setup guide for our Python implementation can be found in GitHub.[1]

---

[1] https://github.com/mallouzi1/Soter

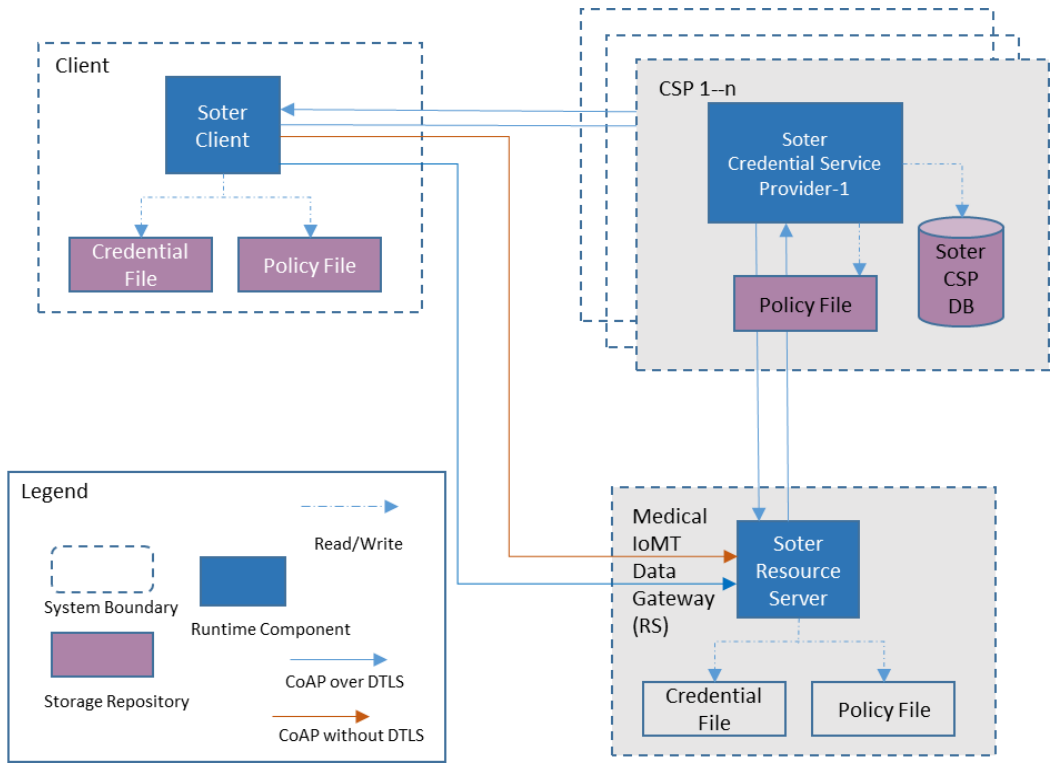

*Figure 6 Soter Dynamic View*

## VII. EVALUATION

### A. Resource Consumption

We evaluated our implementation using the GENI [9] testbed. Figure 7 shows the physical view of Soter, and Figure 8 shows the experiment implemented in GENI. The client requested resources from this experiment's resource server (RS). The access control policy contains credentials from two credential service providers (IdP$_1$ and IdP$_2$. The UDP traffic, Memory Usage, and CPU consumption were measured for the client, RS, and the two IdPs, and the results in Figure 9 show the resource consumption results for the client node. The results are also tabulated in Table III below:

Table III Resource Consumption

| Device | UDP traffic (Packets) | CPU Usage (%) | Memory Usage (MB) |
|---|---|---|---|
| Client | 30 | 19% | 1.15 |
| Resource Server | 25 | 15% | 1.175 |
| $IdP_1$ | 25 | 12% | 1.16 |
| IdP2 | 29 | 13% | 1.16 |

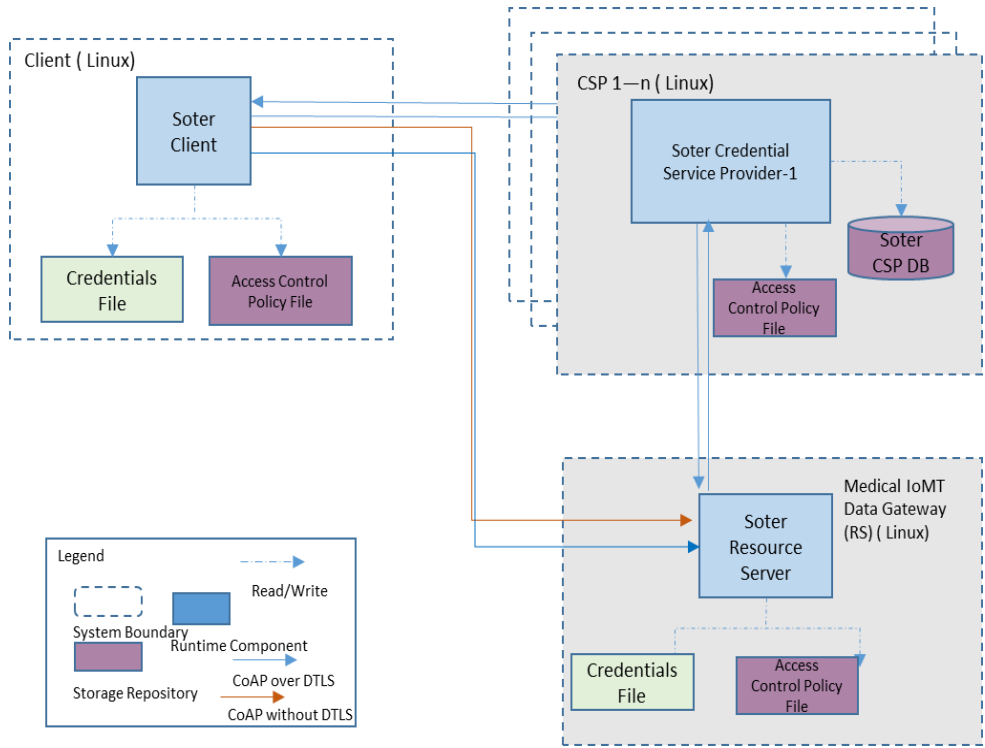

Figure 7 Physical View of Soter

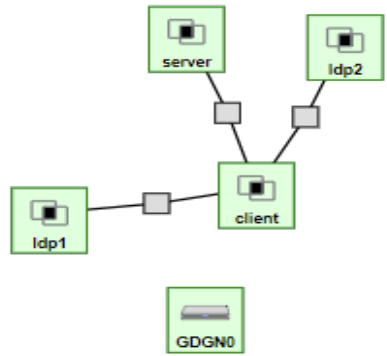

*Figure 8 GENI implementation of $C_1$ $I_2$ $R_1$ model*

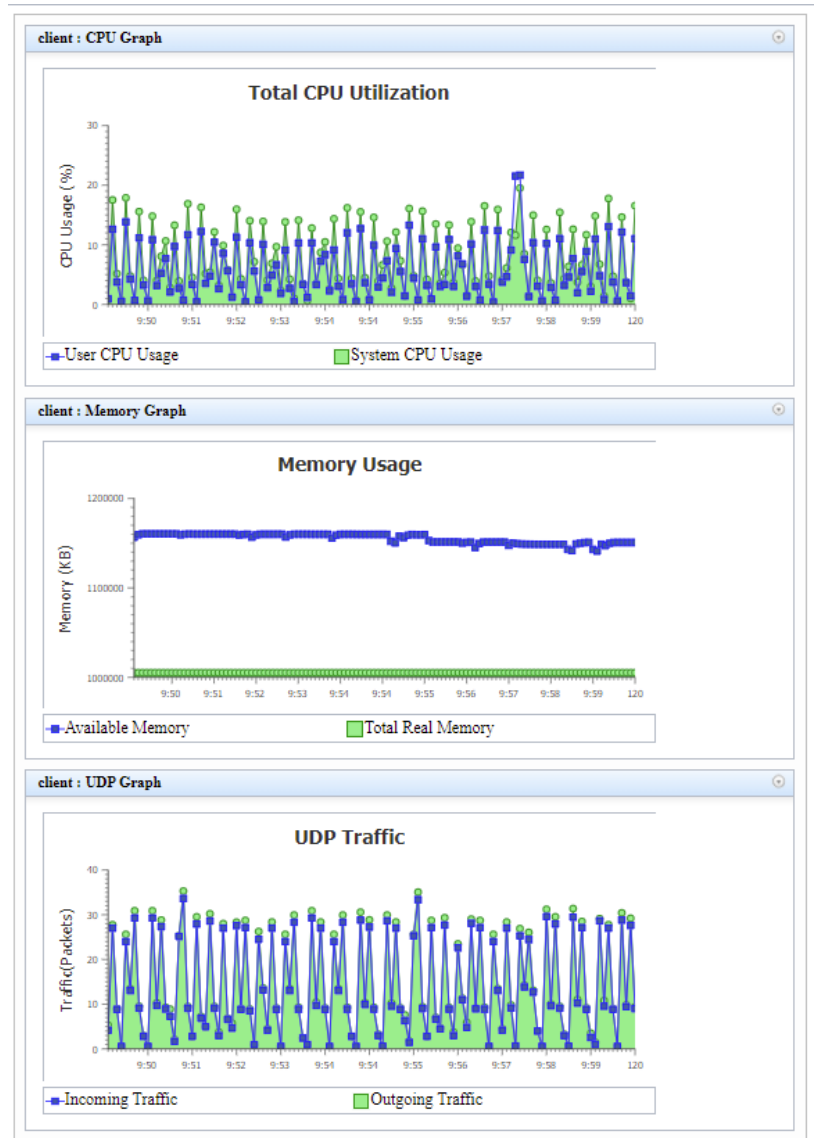

*Figure 9 Resource Consumption at the Client*

### B. Vulnerability Analysis

We need a way to model threats against the Soter system. If we can understand all the different ways Soter can be attacked, we can likely design countermeasures to prevent those attacks. And if we know who the attackers are- not to mention their abilities, motivations, and goals- we can install the proper countermeasures to deal with the real threats.

We perform vulnerability analysis utilizing attack trees [10] to define potential attack vectors. We identified four main attack vectors: a compromised client, Resource server, credential provider, or network. The resulting attack trees are shown in Figure 10-13. The compromised client could lead to unauthorized access to medical records on the medical resource server and access to client credentials on the credential service provider. A compromised credential service provider could lead to unauthorized access to medical records and to tampering with and changing data provided to the clients. A rascal node in the IoMT network could lead to unauthorized access and tampering with medical records. To mitigate these attacks, we proposed the following:

1. Using the COSE protocol will disable the ability to spoof the resource server, as the node would not have a key to decrypt the COSE-encrypted messages.
2. The DTLS protocol prevents man-in-the-middle attacks.

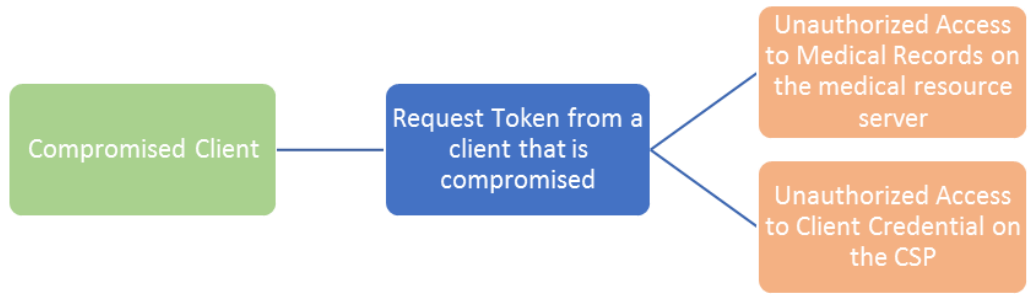

*Figure 10 Vulnerability Analysis-Attack tree 1 : Compromised Client*

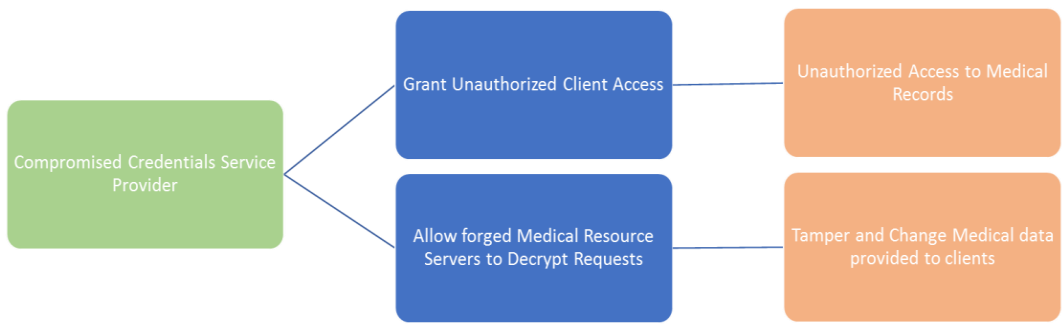

*Figure 11 Vulnerability Analysis-Attack tree 2 : Compromised CSP*

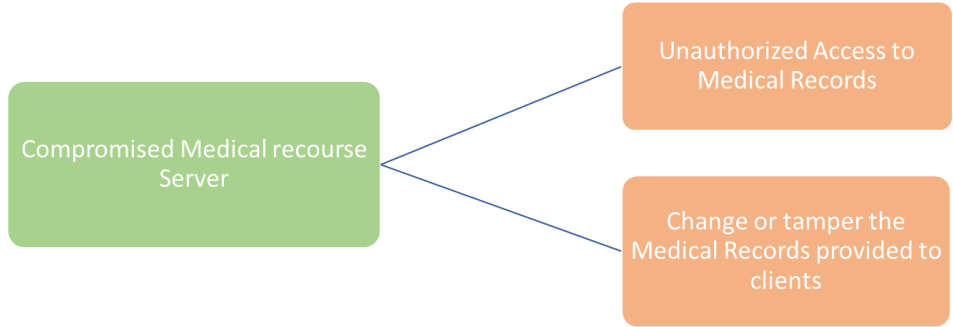

*Figure 12 Vulnerability Analysis-Attack Tree 3: compromised RS*

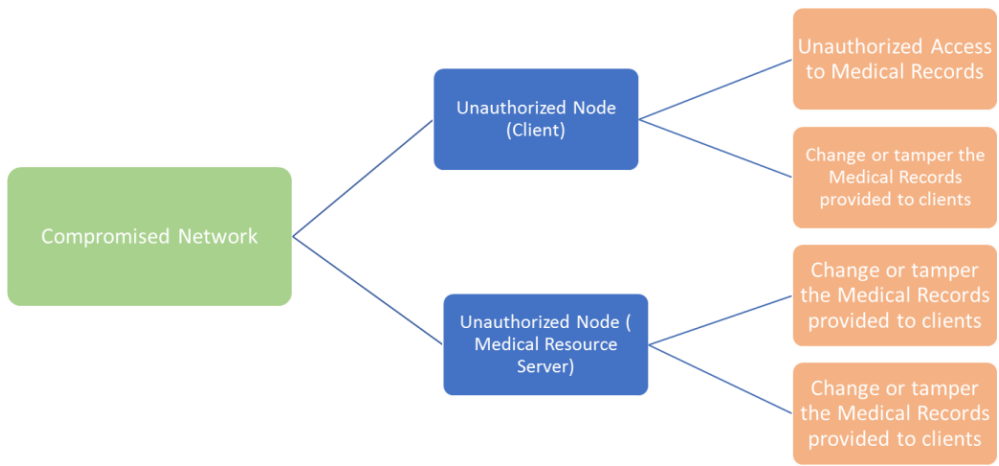

*Figure 13 Vulnerability Analysis-Attack Tree 3: compromised Network*

## VIII. CONCLUSION AND FUTURE WORK

In this paper, we presented an implementation of SOTER, an authentication and authorization system for IoMT-based Zero Trust Management. Threat modeling was conducted as part of the design and architecture process. Resource consumption data and vulnerability analysis were performed as part of the evaluation. The next step for our work includes further optimizing the constrained implementation for Class 1 devices by replacing the encryption protocol with the NIST lightweight authenticated encryption protocol ASCON [11].